\documentclass[conf]{new-aiaa}
\usepackage[utf8]{inputenc}
\usepackage{algorithm}
\usepackage{algpseudocode}
\usepackage{graphicx}
\usepackage{subcaption}
\usepackage{amsmath}
\usepackage[version=4]{mhchem}
\usepackage{siunitx}
\usepackage{longtable,tabularx}
\usepackage{booktabs}
\title{A GPU-Accelerated Sharp-Interface Immersed Boundary Solver for Large-Scale Flow Simulations}

\author{Sushrut Kumar\footnote{Graduate Student.}, Jung-Hee Seo\footnote{Associate Research Professor.} and Rajat Mittal.\footnote{Professor, AIAA Associate Fellow.}}
\affil{Department of Mechanical Engineering, Johns Hopkins University, MD, USA}
\author{Joshua Romero\footnote{Senior Software Engineer} and Massimiliano Fatica.\footnote{Senior Director}}
\affil{NVIDIA Corporation, CA, USA}
\begin{document}

\maketitle

\begin{abstract}
Immersed boundary methods (IBMs) facilitate the simulation of flows around stationary, moving, and deforming bodies on Cartesian grids. However, extending these simulations to the large grid sizes required for realistic flow problems remains a significant computational challenge. In this work, we present the implementation and acceleration of \emph{ViCar3D}, a sharp-interface immersed boundary solver, on graphical processing units (GPUs). We utilize OpenACC, CUDA Fortran and MPI to reprogram \emph{ViCar3D}, a sharp-interface immersed boundary solver, on multi-GPU architectures. Verification and scalability studies are performed for two benchmark cases: two-dimensional flow past a circular cylinder and direct numerical simulation (DNS) of flow past a finite rectangular wing. For the latter, we observe an approximately 20$\times$ speedup (node-to-node comparison) relative to the CPU-based implementation.  The GPU-accelerated solver is capable of simulating complex 3D flows with up to 200 million mesh points on a single node equipped with four GPUs. Strong and weak scaling tests demonstrate maximum scaling efficiencies of 92\% and 93\%, respectively, on multi-GPU systems. We further test the code to simulate fluid flow past complex shaped single body and multi-body cases. 
\end{abstract}



\section{Introduction}
\lettrine{C}{omputational} fluid dynamics (CFD) has significantly advanced research and development in diverse fields such as aerospace, energy, biology and medicine. Accelerating CFD simulations is crucial for enabling high-fidelity simulations of complex internal and external flows and this can be achieved through the development of efficient numerical algorithms and by leveraging modern high-performance computing (HPC) architectures.

On the algorithmic side, a significant focus has been on body-fitted grid methods, where the computational domain is discretized to conform to the geometry of the body. These methods require constructing a body-conforming mesh over which the governing equations—typically representing fluid dynamics and related phenomena like heat transfer—are discretized and assembled into coupled systems of equations. Mesh generation is often a non-trivial task that introduces significant complexity, computational overhead and ``human touch''. These challenges become more pronounced when applying high-fidelity methods such as large eddy simulation (LES) or direct numerical simulation (DNS), particularly for geometrically complex bodies. The computational complexity and burden becomes unsustainable in flow scenarios involving moving or deforming bodies—such as flapping wings, flexible structures, or biological systems like heart valves \cite{mittal2023origin,viola2022fsei,viola2023gpu}.

To address these challenges various immersed boundary methods where simulations are performed on stationary Cartesian grids have been developed, and these are broadly classified into diffuse interface and sharp-interface methods. Diffuse interface methods distribute the influence of the immersed boundary over a neighborhood, which can adversely affect near-wall flow resolution and accuracy due to smeared boundary layers and inexact boundary condition enforcement \cite{mittal2023origin}. In contrast, sharp-interface methods offer a more precise representation of the immersed surface, enabling better resolution of boundary-layer dynamics and more accurate application of boundary conditions. For example, the ghost-cell method, a type of sharp-interface IBM, has been used successfully in simulations of bat aerodynamics, compressible flows, and biological locomotion \cite{mittal2025freeman}.

On the hardware side, most CFD solvers have traditionally relied on central processing units (CPUs) for the bulk of the computational work. Frameworks like the Message Passing Interface (MPI) \cite{walker1996mpi} have allowed simulations to scale across hundreds of CPU cores. However, with the stagnation of Moore’s Law and limits to CPU clock-speed scaling, performance improvements have plateaued—particularly in simulations requiring high Reynolds numbers. In recent years, graphics processing units (GPUs), originally designed for rendering graphics, have gained traction in scientific computing due to their capacity for parallel matrix operations, which are central to both machine learning and CFD. The architecture of GPUs makes them particularly well-suited for accelerating the single-instruction, multiple-data operations common in CFD algorithms, prompting many researchers to port CPU-based CFD codes to GPU platforms \cite{watkins2016multi,lopez2014verification}.

Despite this trend, the adoption of GPUs in the immersed boundary method community has been limited. Notable efforts include the works by Viola et al. \cite{viola2022fsei,viola2023gpu} and Raj et al. \cite{raj2023gpu}. While Viola et al. implement a direct forcing method on GPUs, Raj et al. employ a sharp-interface approach, though applied to relatively simple geometries. There remains a gap in high-performance, GPU-native sharp-interface IBM solvers capable of handling complex unsteady flows relevant to real-world applications.

In this work, we address this gap by developing a GPU-accelerated sharp-interface immersed boundary method designed for DNS of complex unsteady aerodynamic flows. A typical sharp-interface IBM framework comprises two primary components: a geometric module, responsible for representing the immersed boundaries, and a numerical module, which solves the governing partial differential equations. Here, we focus exclusively on the numerical module, which is broadly applicable across problems involving both stationary and moving boundaries. Porting and GPU-acceleration of the geometric module is reserved for future work.

\section{Computational Methods}
In this section, we will elaborate on the numerical methodology and considerations for developing the GPU-accelerated flow solver.
\subsection{Governing Equations and Discretization Methods}
In our work, we perform direct numerical simulation of incompressible flow by solving the conservative form of Navier-Stokes in 3D and time, as shown in the equation \ref{eq:NS}.
\begin{equation}
    \frac{\partial u_i}{\partial t} + \frac{\partial u_i u_j}{\partial x_j} = - \frac{\partial p}{\partial x_j} +  \frac{1}{\text{Re}}\frac{\partial^2 u_i}{\partial x_j^2}; \, \, \, \frac{\partial u_i}{\partial x_i}=0 
    \label{eq:NS}
\end{equation}
We utilize the fractional step method\cite{CHORIN196712} to split equation \ref{eq:NS} into an advection diffusion equation and a pressure Poisson equation.
We utilize the central difference scheme for spatial discretizations and Adams-Bashforth for temporal discretization. Once the equations are discretized these have to be represented over a computational mesh which in our case is an non-uniform Cartesian grid as shown in figure \ref{fig:meshes}. Then the equation are assembled to a linear sparse system of form, ($Ax=B$) and solved using a iterative method to obtain the solution vectors. Different methods are available to solve this equation such as line successive-over-relaxation (LSOR), red-black Gauss Seidel, multigrid methods, and Krylov subspace methods. GPUs execute tasks using the SIMT (single instruction multiple threads) model and methods that allow code to be written in single instruction format would result in massive speedups. We have evaluated various methods such as LSOR, multigrid methods \cite{menon2022contribution} and biconjugate gradient stabilized method (BiCGStab) \cite{kumar2025computational} for pressure Poisson solver. So far we have obtained the best performance with BiCGTAB and proceed forward with this method. We also utilize a scheduled relaxed Jacobi solver \cite{YANG2014695} for solving the advection diffusion equation and as a preconditioner to BiCGStab solver. Both these choices are appropriate for GPU programming model. 

Additionally, the sharp representation of Dirac-delta type discontinuity in the flow that exists would require additional treatment and a scalable method require the algorithmic treatment to be of SIMT type as well. Typically, the logic to implement these algorithms would require multiple conditional statements along with scalar value reductions which can be expensive to perform and performing these operations on GPU could give rise to issue such as warp divergence and race conditions. We will elaborate more on the implementation later.

\begin{figure}[hbt!]
\centering
\includegraphics[width=1\textwidth]{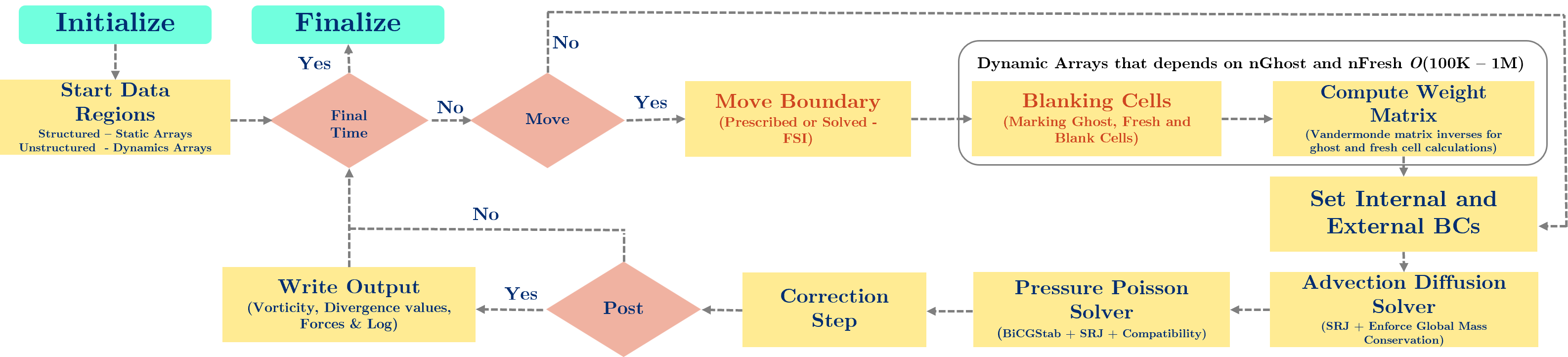}
\caption{Data management and GPU offloading}
\label{fig:flowChart}
\end{figure}

We present the process flow in figure \ref{fig:flowChart}. Here, an important step after initialization is to introduce structured (fluid volume) and unstructured (surface description) data regions. The structured data regions are used for primary arrays that do not change shape during the simulations, whereas, the unstructured data regions are for dynamic array connected with the immersed body geometry, and these may in principle change for deforming bodies. The purpose of prescribing these data regions is to inhibit any expensive host to device and device to host data movements during the simulation. After declaring the start of data regions, we perform the usual time-stepping procedure by first moving the immersed body (either prescribed or through fluid-structure interaction \cite{kumar2025computational,kumar2025mechanical}), then applying the internal and external boundary conditions followed by solving advection diffusion and pressure Poisson equation. This procedure is standard for most CFD codes but the key to achieving speedup on GPUs is to utilize CUDA streams and asynchronous launches where ever possible to minimize the kernel launch overhead.  

\subsection{Treatment of Immersed Body}
\begin{figure}[hbt!]
  \centering
  \begin{subfigure}[b]{0.4\textwidth}
    \centering
    \includegraphics[width=\textwidth]{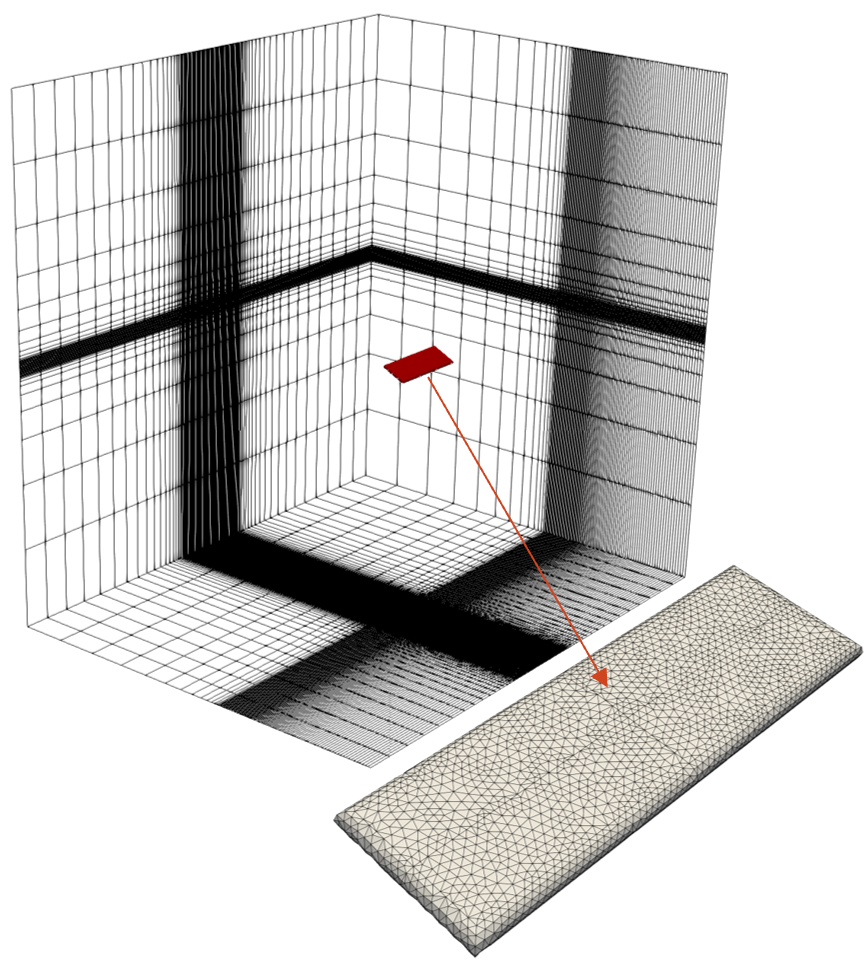}
    \caption{}
    \label{fig:meshes}
  \end{subfigure}  
  \hspace{1cm}
  \begin{subfigure}[b]{0.35\textwidth}
    \centering
    \includegraphics[width=\textwidth]{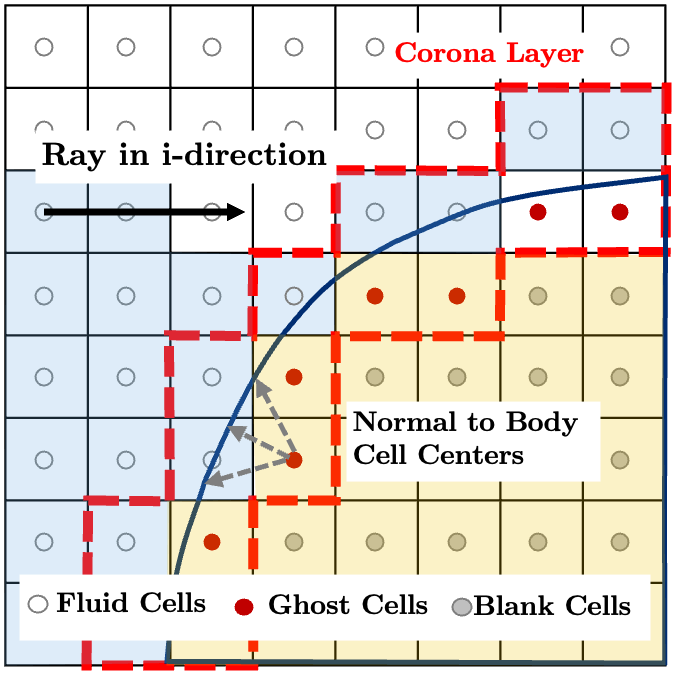}
    \caption{}
    \label{fig:iblank}
  \end{subfigure}

  \vspace{1em}

  \begin{subfigure}[b]{0.35\textwidth}
    \centering
    \includegraphics[width=\textwidth]{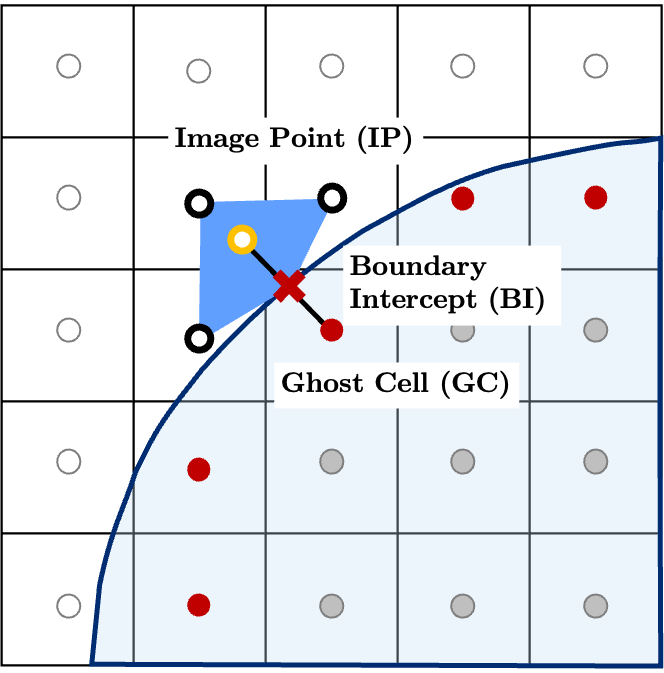}
    \caption{}
    \label{fig:GC}
  \end{subfigure}
  \hspace{3cm}
  \begin{subfigure}[b]{0.35\textwidth}
    \centering
    \includegraphics[width=\textwidth]{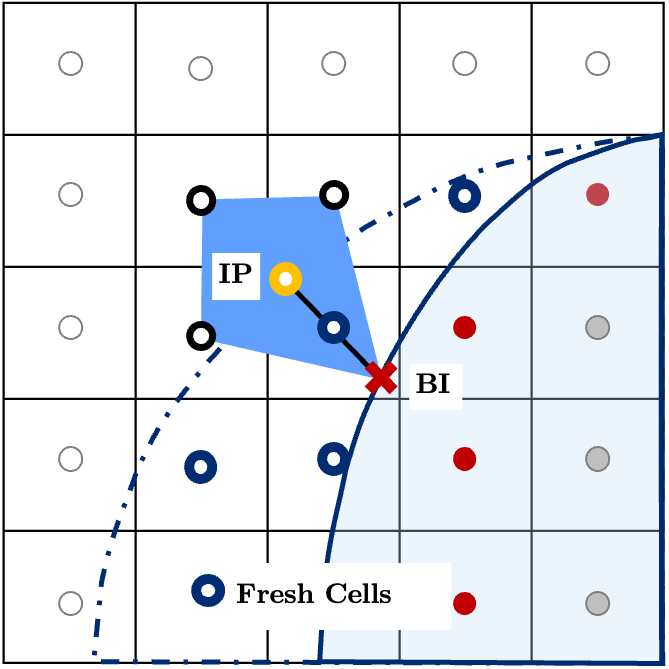}
    \caption{}
    \label{fig:FC}
  \end{subfigure}

  \caption{Treatment of Sharp-Interface Immersed Boundary within Solver. (a) Cartesian grid for flow fields with immersed body represented using triangulated mesh. (b) Ray tracing procedure to mark different cells. (c) Calculation of ghost cell values. (d) Calculation of fresh cell values.   }
  \label{fig:2x2grid}
\end{figure}

In this work, we utilize the ghost-cell method (GCM), a variant of sharp-interface immersed boundary methods as proposed in Mittal et al. \cite{mittal2008versatile,mittal2025freeman}. The application of GCM first requires classification of cells within the computational domains into fluid, ghost, fresh and dead cells. The "fluid cells" are the ones where the flow is present and governing equations are solved. The "dead cells" are the cells present within the immersed body and no equation need be solved here. The "ghost cells" are dead cells with fluid cell neighbor and are  cells to used to implement Dirichlet and Neumann boundary conditions on the body. In case of moving bodies, dead cells transition into fluid cells which are termed as "fresh cells" and they require special treatment as no information is usually present to estimate derivatives for the governing equation. At it's core, GCM takes into account the curvature of body and specific boundary condition to populate ghost and fresh cells. The curvature information is provided to the code using the unstructured surface mesh and this mesh is always immersed within a Cartesian grid as shown in figure \ref{fig:meshes}. An advantage of this step over body-fitted grid is that it eliminates the expense of generating a new body-fitted computational grid at every timestep for a moving/deforming, and reassembling the linear systems for new grid topologies.

The process of incorporating immersed body motion and application of boundary conditions consists of three sub processes: (a) locating and blanking of the dead (or solid) cells within the body (see figure \ref{fig:iblank}), (b) calculating quantities at the ghost cells "$\phi_{GC}$" (see figure \ref{fig:GC}) and (c) calculating quantities at fresh cells "$\phi_{FC}$" (see figure  \ref{fig:FC}). We utilize a ray tracing algorithm to demarcate the fluid cells and solid cells. The ray tracing algorithm is ubiquitous in computer graphics and can be easily optimized for GPUs, see \cite{ruetsch2024cuda}. Once the fluid and dead cells are demarcated, we can then proceed to classify fresh and ghost cells. This step is followed by finding the closest points on the surface of immersed body from ghost cell where a normal can be dropped. This points is termed as "body-intercept (BI)" in our work, see figure \ref{fig:GC}. Once the BI is known, we extend the normal into the flow domain and identify a ``image point'' (IP). The next step is to find 8 fluid cells surrounding the IP. We can then compute the quantities at IP using 
\begin{equation}
    \phi_{IP} = \sum_{i=1}^{8}\beta_i \phi_i
\end{equation}
Here, $\beta_i$ is the interpolation constant of each one of eight neighbouring fluid cells. The interpolation constant is a function of spatial coordinate of IP $\Vec{x}_{IP}$ and another vector $C$ computed as 
\begin{equation}
    C = [V]^{-1}\phi;
\end{equation}
$V \in \mathbb{R}^{8\times8 \times n_G}$ is the Vandermonde matrix whose inverse is computed for all the image points surrounded by 8 fluid cells. The process of computing $C$ requires inverting a very large number of small $8\times8$ matrices. We compute the inverse of these matrices by using functions \texttt{cublasDgetrfBatched} and \texttt{cublasDgetriBatched} available within the cuBLAS library with batch size equal to number of ghost cells $n_G$. Once the quantity of interest, is computed at the image-point, we can then utilize the central difference scheme along with Dirichlet and Neumann boundary conditions to obtain implicit equations that reside in our linear system. A similar process can be followed to obtain the implicit equation for the fresh cell with an exception of accounting for body velocities evaluated at BI as shown in figure \ref{fig:FC}.

\subsection{Multi-GPU Implementation}
\begin{figure}[hbt!]
  \centering
  \begin{subfigure}[b]{0.45\textwidth}
    \centering
    \includegraphics[width=\textwidth]{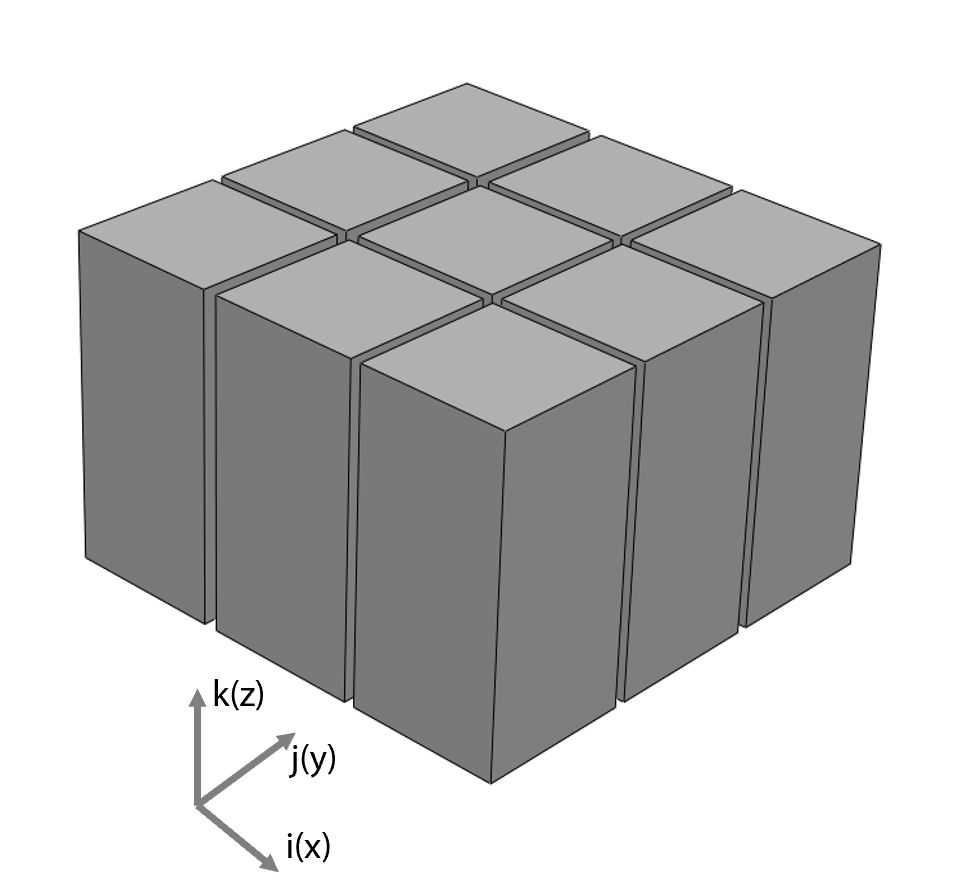}
    \caption{}
    \label{fig:domain-decomp}
  \end{subfigure}
  \quad
  \begin{subfigure}[b]{0.45\textwidth}
    \centering
    \includegraphics[width=\textwidth]{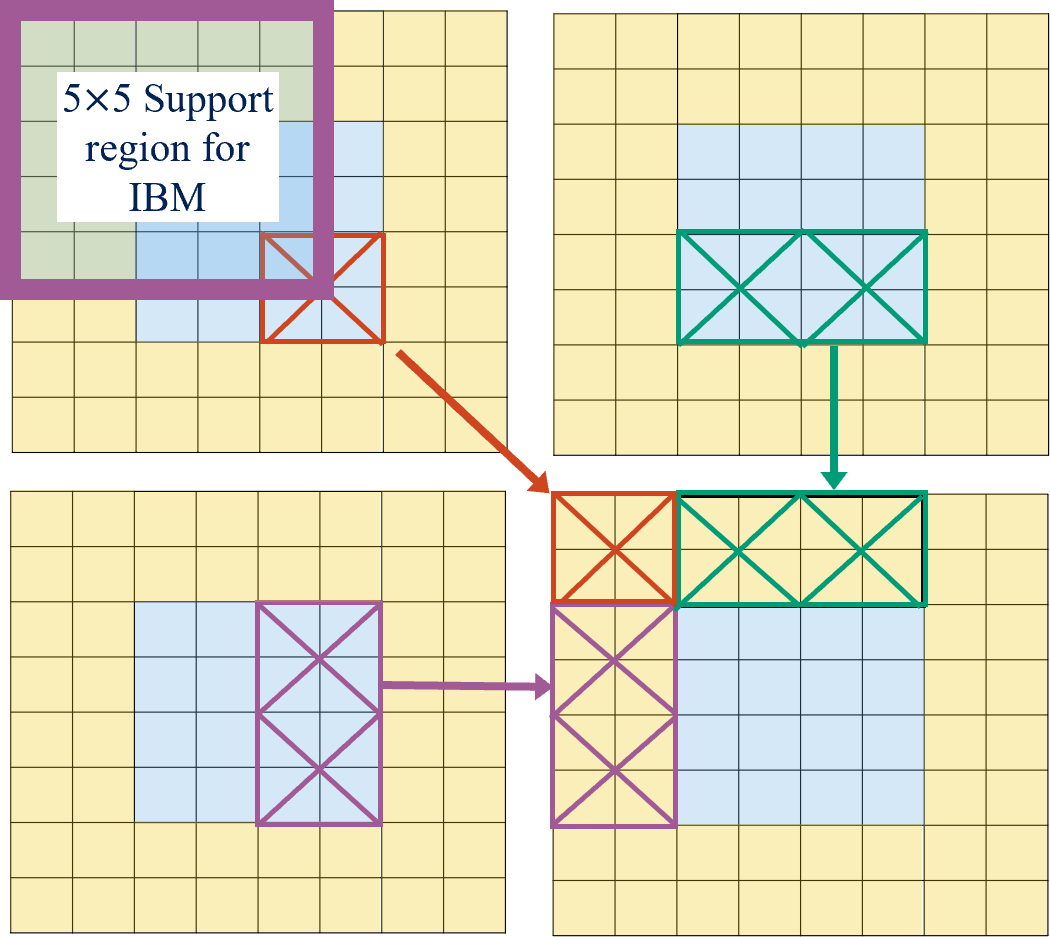}
    \caption{}
    \label{fig:mpi}
  \end{subfigure}
  \caption{Schematic for domain decomposition. (a) pencil like domain decomposition of cartesian indices. (b) Communication of data between neighboring partitions}
  \label{fig:domain}
\end{figure}

GPU systems are extremely fast but can be bottle-necked by the limited device memory used for storing arrays during the simulation. As mentioned earlier, as our solver is fully GPU-native, it becomes a even bigger challenge as it demands transferring all the arrays before the start of time-stepping procedure. This challenge is alleviated by using communication across multiple GPU devices in a manner similar to CPU communication in multi-CPU CFD software. Another advantage of employing a Cartesian grid is that it is trivial to partition into the pencil-like domain using 2D Cartesian decomposition as shown in figure \ref{fig:domain-decomp}. The partitioning is done in the index space and not in Cartesian coordinate space. Further, this decomposition ensures that the computational load is always balanced across partitions even when the body undergoes large displacements. This eliminates the need for complex methods like graph partitioning popularly used in body-fitted grid methods as it will change the communication topology every timestep when body makes large displacements in order to balance the computation load. Moreover, having pencil-like partitions is advantageous when using finite-difference methods as it becomes trivial to perform halo exchanges and access neighboring partition data.

We utilize GPU-aware MPI and "partial-blocking" methodology to communicate across GPUs. GPU-aware MPI is necessary to eliminate any expenses of CPU staging of data that will occur before communication. The cost is exceptionally high when communicating large halo regions for problems with high Reynolds numbers ($Re$) which requires a high number of grid points to resolve flow structures. Next, the GCM necessitates the availability of a $5\times5$ grid point support region, as shown in figure \ref{fig:mpi}, due to the curvature of the body to apply the internal boundary conditions and compute flow variables at fresh cell. This requires us to properly communicate corner regions along with horizontal and vertical direction to accurately solve the system.

A common method to communicate between different neighboring ranks is to use blocking method through the "\texttt{MPI\_sendrecv}" command. This method blocks the computation along with communication that can happen at other faces. We eliminate this by using a two stage communication procedure referred as "partial-blocking". In our implementation, we first communicate the data between the partitions in the north and south direction followed by east and west directions where each communication will have two data sends and two data receives. Both the movements are facilitated by \texttt{MPI\_Isend} and \texttt{MPI\_Irecv} commands. Further, this two-step communication ensures proper communication of corner halo regions required for our $5 \times 5$ support region. With this approach, even in Cartesian grid we cannot always guarantee that the data to communicate is contiguous in memory and in-fact in pencil like domain decomposition it is mostly non-contiguous. In a CPU code, this issue is resolved using MPI derived datatype such as \texttt{MPI\_Type\_vector} but we have found that using this is almost $O(10^2)$ more expensive than directly communicating contiguous memory using GPU-aware MPI. Next, we have found that using contiguous buffer arrays which are packed by launching CUDA kernels is an efficient way of sending non-contiguous arrays. The wall-time even with kernel launch for packing buffer was the same order-of-magnitude as directly sending contiguous data.

\section{Results and Discussion}
 This section presents the scaling, verification test of our developed code. The tests have been performed on two different computing platforms: JHU ARCH Rockfish and DoD AFRL Raider. We first present the performance scaling on multiple GPUs followed by verification of the GPU solver using two tests - 2D flow past a circular cylinder and 3D flow over a rectangular wing, and by comparing against published results from simulations done using the CPU code. Next, we scale the rectangular wing simulation to a  high Reynolds number and a much larger grid to demonstrate the code's capability to perform direct numerical simulations of such systems. All of these tests have so far been done using single node L40s (on Rockfish) and  A100 (Raider). Each L40s node has 8 GPUs connected with PCIe and each A100 node has 4 GPUs connected with NVLink. Further hardware details will be given in subsequent sections. The code is compiled using the NVHPC SDK 23.1 and uses in-built openMPI implementation for parallelization.

\subsection{Performance Comparison}
\begin{figure}[hbt!]
  \centering
  \begin{subfigure}[b]{0.47\textwidth}
    \centering
    \includegraphics[width=\textwidth]{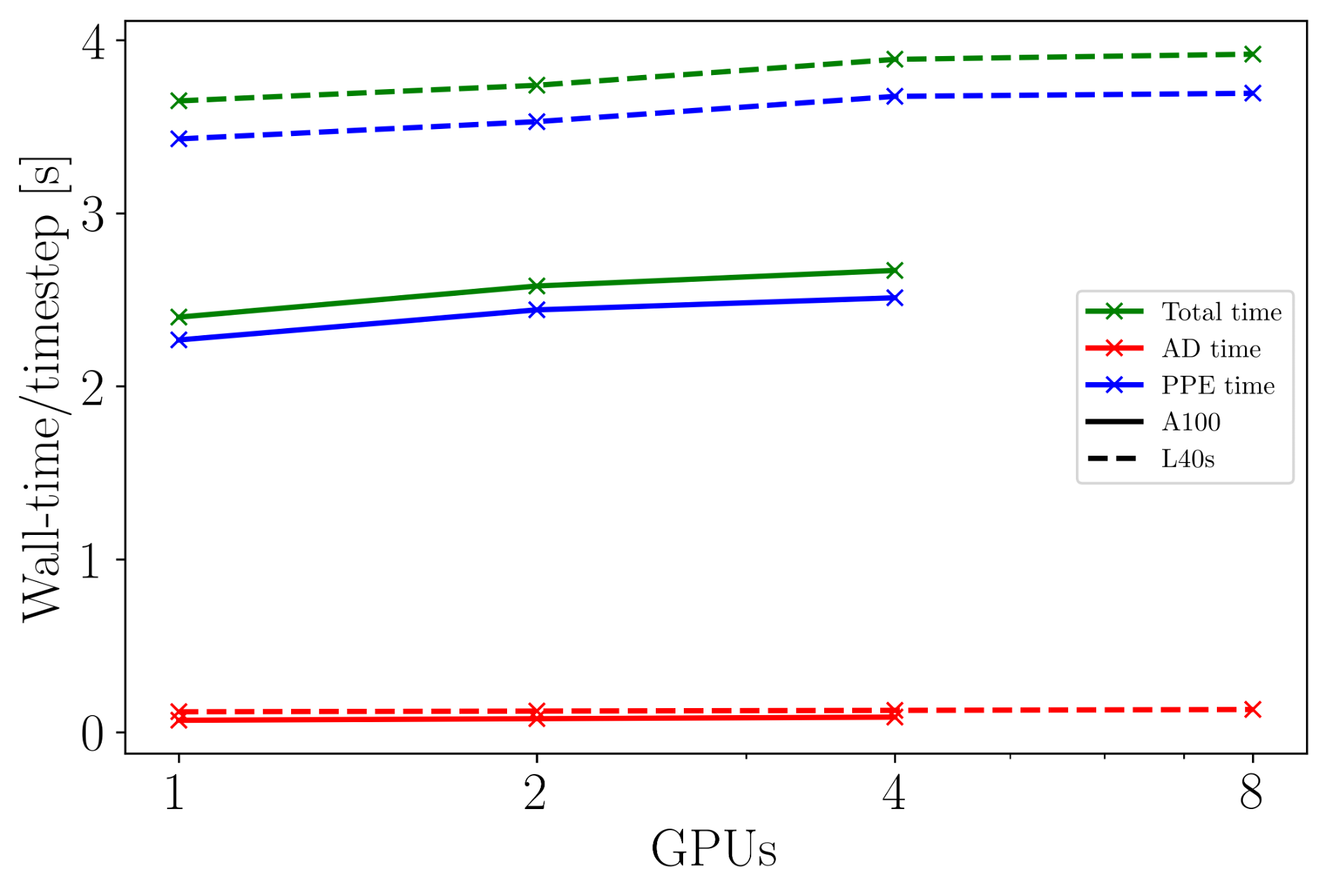}
    \caption{}
    \label{fig:weak}
  \end{subfigure}
  \quad
  \begin{subfigure}[b]{0.49\textwidth}
    \centering
    \includegraphics[width=\textwidth]{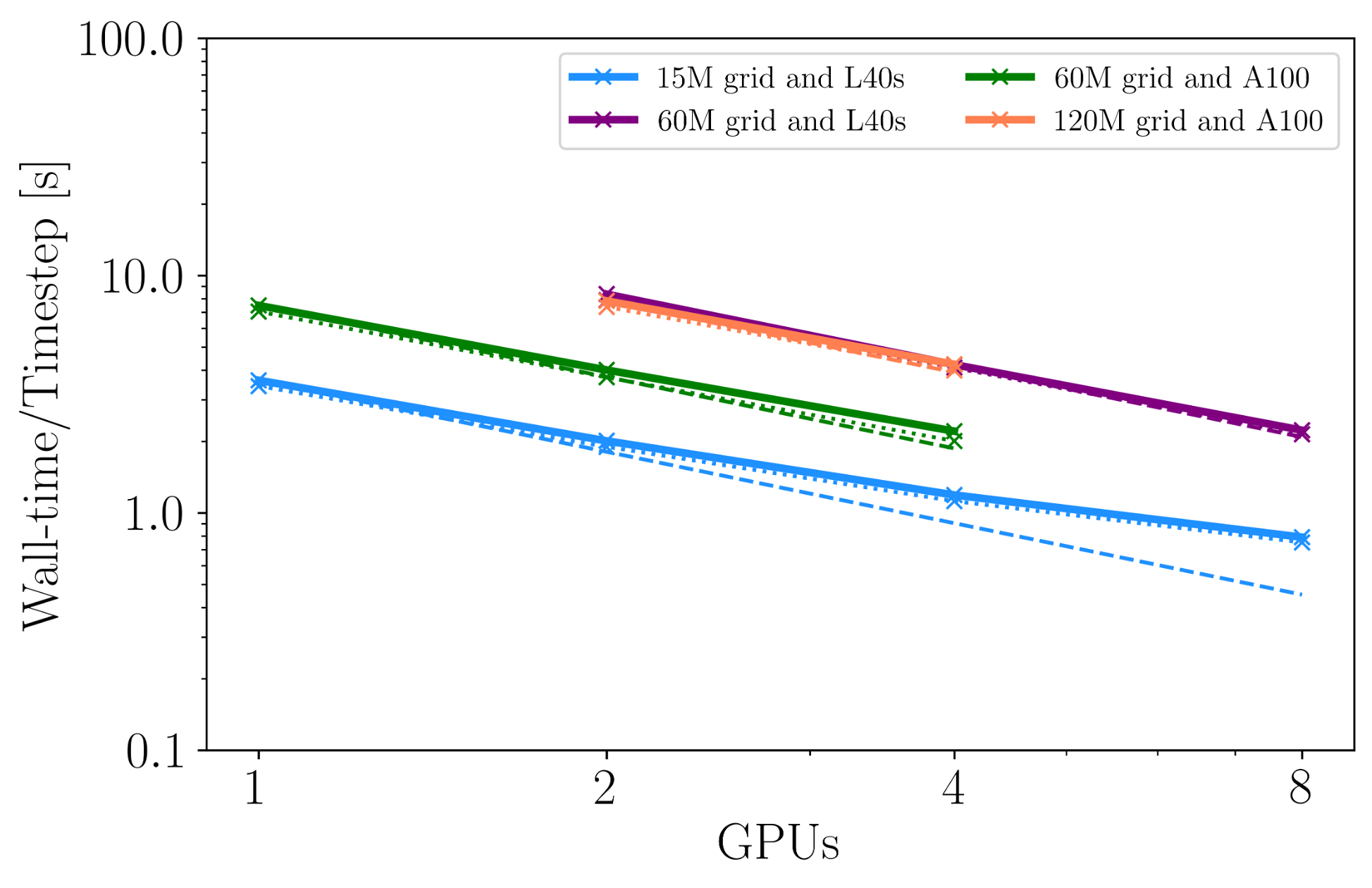}
    \caption{}
    \label{fig:strong}
  \end{subfigure}
  \caption{Scaling tests. (a) Weak scaling test. (b) Strong scaling test}
  \label{fig:domain}
\end{figure}

In this section we present the solver's capability to scale to multiple GPU devices. We employ two different tests - weak scaling and strong scaling. Figure \ref{fig:weak} and \ref{fig:strong} presents wall clock time for weak and strong scaling tests respectively. We perform the weak scaling by assigning a fixed per-GPU subdomain resolution of $301 \times 385 \times 129$ grid points in three spatial direction which is about 15 million (M) grid point for CFD of flow around a 3D rectangular wing at $Re=1000$ using double precision. We then increase the global mesh proportionally up to 4 and 8 GPUs on A100 and L40s platforms respectively. On the A100 platform each with 40 GB of memory per GPU, the wall-time per time step starts at 2.40s per GPU and remain constant as the GPUs are increase to 2 GPU with wall-time 2.58s and 3.67s for 4 GPUs. This yields a weak scaling efficiency of about 90\%. On the L40s, the wall-time starts at 3.65s and goes up to 3.92s indicating an efficiency of about 93\%. On both the platforms, the slight increase in wall-time can be attributed to increase in communication between nodes. We would also like to mention that even though we use asynchronous streams to launch our CUDA kernels, the streams have to be blocked before communication as MPI is not stream-aware. This adds a slight overhead for synchronization since the GPUs are idle after the packing and unpacking of buffer arrays. 

Figure \ref{fig:weak} also highlights the wall-time fraction for different element of the solver such as the pressure Poisson solver, advection-diffusion solver and other elements for stationary body simulations. It can be seen that the pressure Poisson is the most expensive element of the solver. Moreover, we have found A100 to be faster for our calculation as it has dedicated FP64 support and offers a higher memory bandwidth than L40s, which only offers FP32 support. These tests demonstrate our software ability to scale  with the number of GPUs as the problem size increases.

Next, we performed a strong-scaling study to evaluate the solver's ability to accelerate a given problem as the number of GPUs increases. Tests were conducted using three baseline grid sizes---15M ($301\times385\times129$), 60M ($603\times771\times129$), and 120M ($1205\times771\times129$)---to probe the latency, throughput, and communication limits of the implementation.
Because each GPU kernel launch incurs a fixed overhead, smaller grids experience proportionally higher launch costs relative to computation. This is evident in the 15M case, where scaling efficiency drops to approximately 60\% at larger GPU counts; for example, the 8-GPU configuration corresponds to only $\sim 200{,}000$ grid points per GPU.
In contrast, the 60M grid exhibits substantially better scaling, achieving
efficiencies of 86\% on A100 GPUs and 90\% on L40s. For the 120M grid, the full
domain does not fit on a single GPU, so we report results only for 2- and 4-GPU
configurations, where the solver maintains a normalized efficiency of 92\%.
Overall, these results demonstrate excellent strong-scaling performance,
particularly for large grids, and highlight the importance of selecting
sufficiently large per-GPU workloads to fully exploit modern GPU architectures.

\subsection{Verification Cases}
\subsubsection{2D Flow past a Circular Cylinder at $Re=1000$}
\begin{figure}[hbt!]
\centering
\includegraphics[width=1\textwidth]{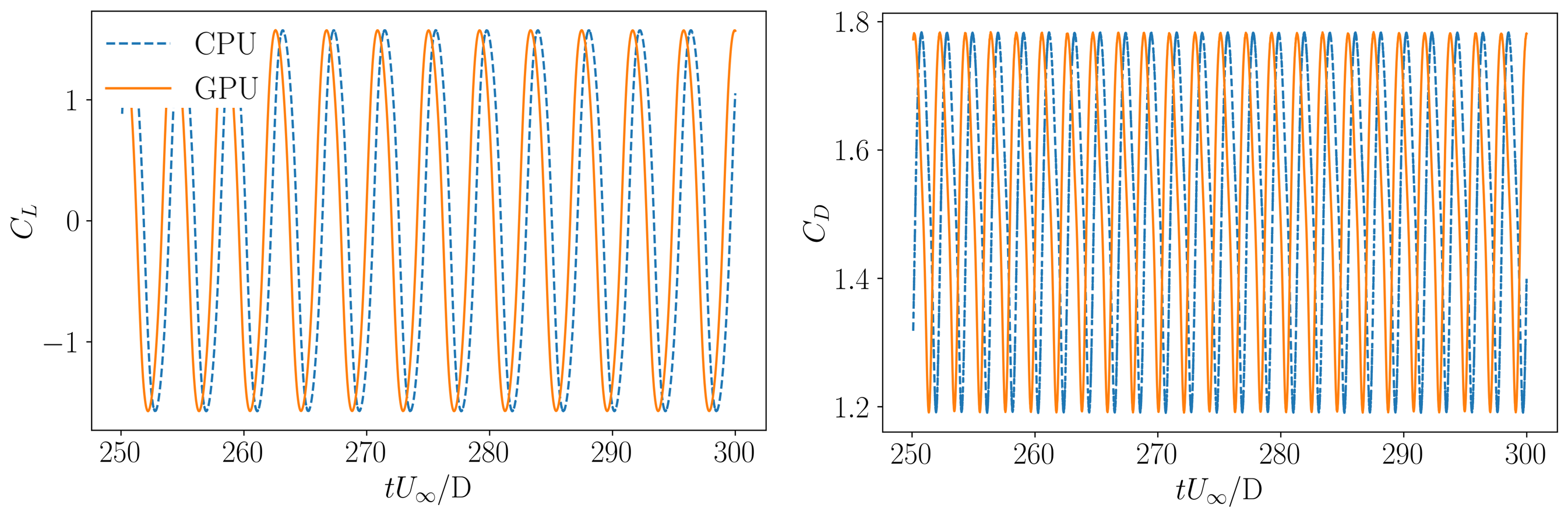}
\caption{Aerodynamic forces for flow over 2D Cylinder at $Re=1000$}
\label{fig:2dcylinder}
\end{figure}

This case a canonical case for code verification and is therefore used as the first case to demonstrate the fidelity of our code. A $301\times385$ Cartesian grid is used for this $Re=1000$ flow and the simulation was done on one A100 GPU using FP64 precision. The time-series data of key aerodynamic quantities of lift, $C_L$ and drag, $C_D$ after the stationary state is reached are presented in figure \ref{fig:2dcylinder}. The phase shift in the two plots is due to the expected differences in initial transient state as the vortex shedding in our simulation is initiated by the exponential growth of roundoff errors, which are different for the GPU and CPU codes, and not because of any prescribed disturbances. We obtained a mean drag coefficient, $\overline{C}_D$ of about 1.51 and this is in perfect agreement with the results published by Mittal et al. \cite{mittal2008versatile} from the CPU version of this code.

\subsubsection{3D Flow over a Rectangular Wing at $Re=1000$}
\begin{figure}[ht]
\centering
\includegraphics[width=1\textwidth]{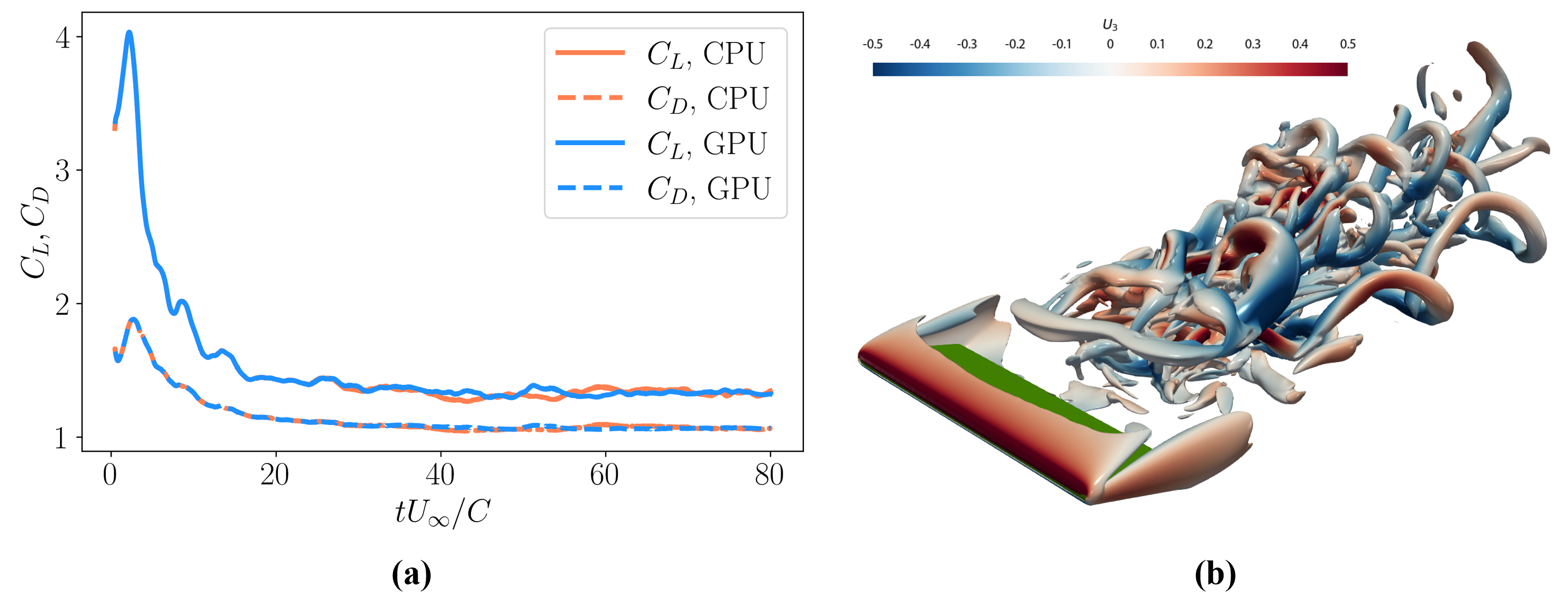}
\caption{Aerodynamic forces and flow structures over the finite 3D wing at $Re=1000$}
\label{fig:3dwing}
\end{figure}

Next, we tested our code for accuracy in simulating 3D flows. The reference case is adopted from Menon et al. \cite{menon2022contribution}.  The Reynolds number is defined based on the wing chord, $C$, and freestream flow velocity $U_\infty$. The computational domain is a 3D Cartesian grid which extends $10C$ in all three directions and shown in figure \ref{fig:domain}. Figure \ref{fig:domain} also shown the triangular mesh of the immersed wing which has a aspect ratio (span to chord) of 3 and is inclined at an angle of attack, $\alpha=25^\circ$. The high angle-of-attack will result in a stochastic flow separation and turbulent-like flow in the wake and is therefore a strong test for the accuracy of the GPU code. The CPU simulation was performed using 10 nodes of Intel Xeon Gold Cascade Lake 6248R, each with 48 cores per node. The simulation took 56 hours to perform 80 convective time units using this 480 cores configuration. The GPU simulation was performed using a single A100 node with 4 GPUs and the simulation were done with FP64 precision and took about 24 hours. The same simulation with the CPU code will have a theoretical wall-time of 560 hrs assuming a 100\% scaling efficiency and this translates to about $20\times$ speedup between a single A100 node and a single Xeon CPU node. The quantitative comparison of aerodynamic force coefficient is presented in figure \ref{fig:3dwing}. It can be seen that we were able to achieve a very good agreement between CPU and GPU results, with the differences between the two a result of the instability growth process that results in the vortex shedding and which results in a separation in the phase trajectory of the two cases, as is expected for a chaotic system such as this one. 
\begin{table}[ht]
  \centering
  \caption{Total wall-time per simulation for different configurations.}
  \label{tab:time_per_timestep}
  \begin{tabular}{@{}lccc@{}}
    \toprule
    Case         & 1 CPU node (48 Cores) & 10 CPU nodes & 1 GPU Node (4 GPUs) \\
    \midrule
    3D wing      & 560 h (theoretical) &  56 h & 25 h \\
    \bottomrule
  \end{tabular}
\end{table}

\subsection{Complex Test Cases}
Here, we focus on testing the code for complex cases after verifying the correctness of the solver to the CPU version of code. In this section, we present complex test cases that involve complex curved bodies to multi-body setup to demonstrate the capability of our sharp-interface immersed boundary method in simulating fluid flow without a complex meshing procedure.
\subsubsection{Flow over a Conceptual Flying Vehicle at $Re = 25,000$}
\begin{figure}[hbt!]
\centering
\includegraphics[width=1\textwidth]{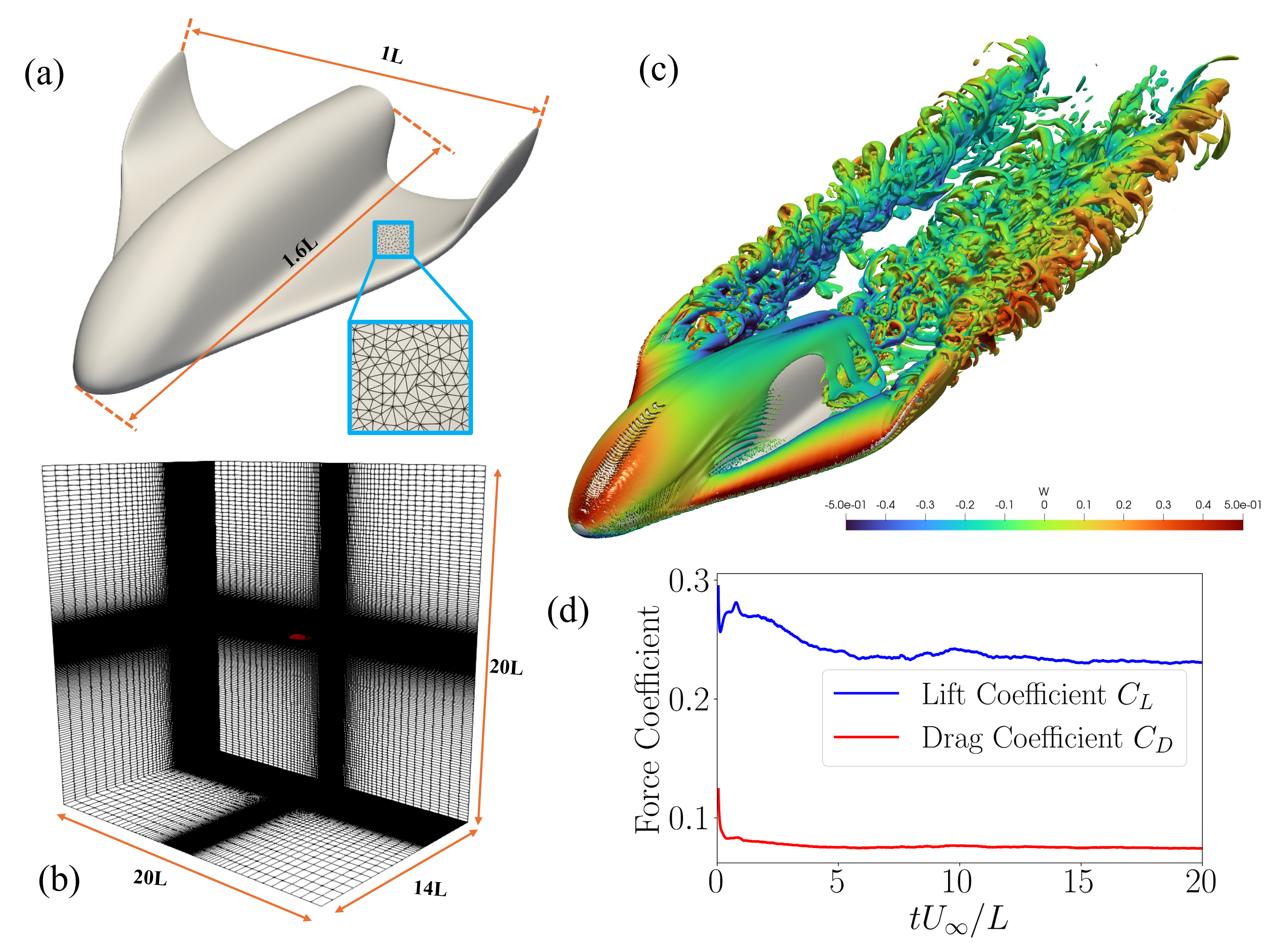}
\caption{Complex-shaped flying object (a) CAD model of geometry, (b) Cartesian grid for flow simulation, (c) Isosurfaces at $Q=20$ colored by vertical velocity, $W$, (b) Time-series data of lift and drag coefficient}
\label{fig:uav}
\end{figure}

The first case chosen is that of a complex-shaped flying object as shown in figure \ref{fig:uav}(a) and sourced from the website \texttt{www.fetchcfd.com}. We set the simulation's length scale to the object's wingspan. The object is given an angle of attack $\alpha=25^\circ$ to allow for complex vortex shedding due to incoming fluid flow interacting with the body. The particular case is interesting because of the varying thickness across the entire geometry, including an ellipsoid-like fuselage and thin wings with curved winglets. A body-conformal mesh for such a body would be relatively complex but here we use the Cartesian grid shown in figure \ref{fig:uav}(b) which has  $800\times500\times484$ points in the three directions and a total of about 194 million grid points. As can be seen in figure \ref{fig:uav}(b), the majority of the grid points are focused around the flying object to resolve the boundary layer and wake structure. Further, the grid is stretched rapidly in vertical and spanwise direction, as the flow gradients in these regions are small. 

The grid and geometry for this case is shown in figure \ref{fig:uav}  and the Reynolds number based on the freestream velocity and wing span is set to $Re = 25000$. The simulation was performed on a single node with four NVIDIA A100 GPUs connected via PCIe. The timestep size was set to $\Delta t = 0.0005$ and the simulation took 35 hours to complete the reported simulation time of $t U\infty/L = 20$. The simulation results are presented in figure \ref{fig:uav} (c) and \ref{fig:uav} (d).

\subsubsection{Flow over an array of ellipsoidal particulate flow at $Re = 10,000$}
\begin{figure}[hbt!]
\centering
\includegraphics[width=1\textwidth]{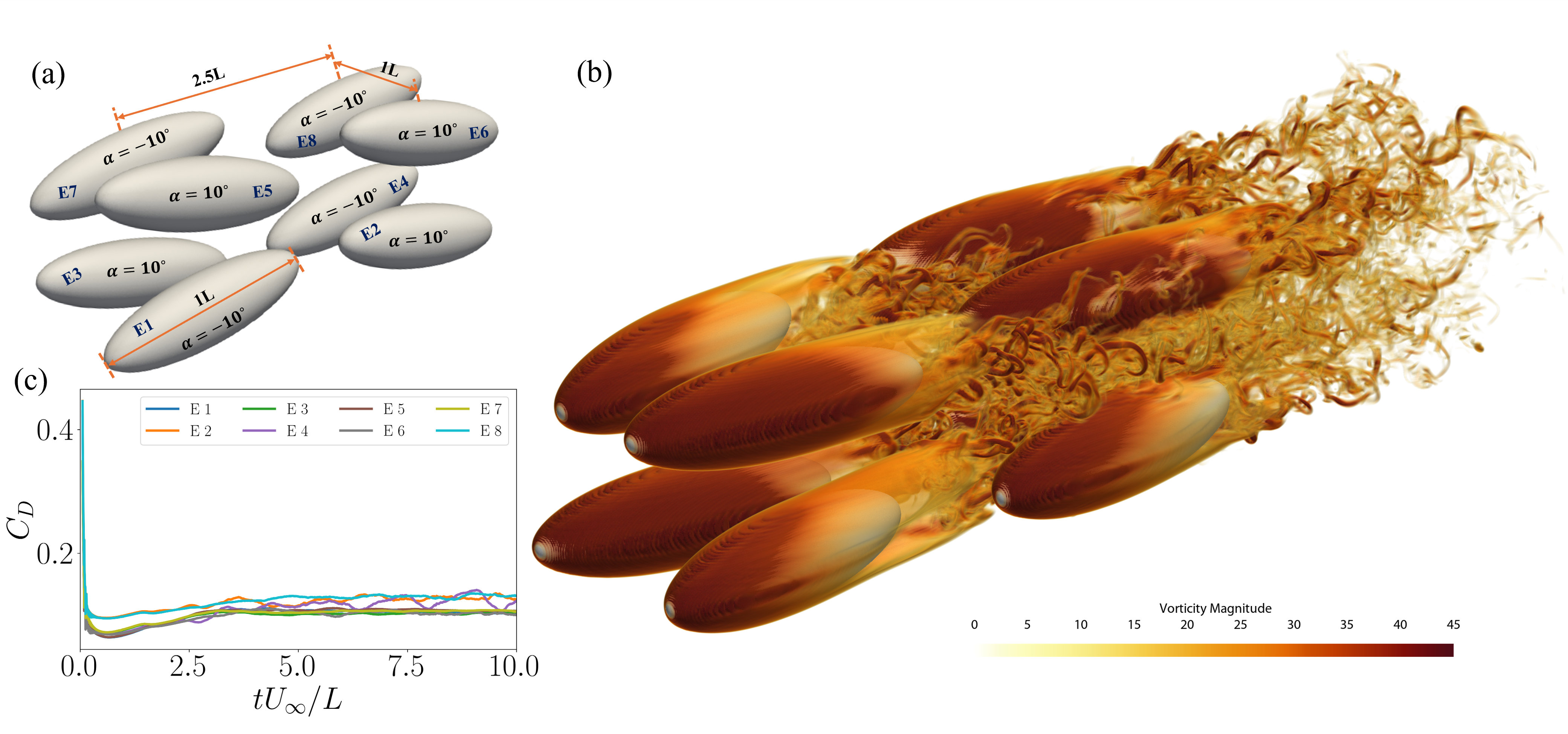}
\caption{Ellipsoidal particulate flow (a) CAD model of geometry, (b) Volumetric rendering of Vorticity magnitude (c) Time-series data of drag coefficient for all bodies}
\label{fig:ell_array}
\end{figure}

After testing the code's efficacy in simulating flow past a complex-shaped object, we shift our focus to simulating fluid flow past a multi-object array. We consider an array of ellipsoids with an aspect ratio of 1(D):0.3. Each ellipsoid is rotated to provide an angle-of-attack, thereby demonstrating two distinct capabilities of our GPU-accelerated IBM method. First, generating the computational body-fitted grid for the arrangement as shown in figure \ref{fig:ell_array}(a) would be highly complex, but it is extremely trivial with a Cartesian grid. The length scale for the simulation is set to the major axis of the ellipsoid and the Reynolds numbers was set to 10,000 and a fine resolution is required to resolve the complex flow through this ellipsoidal array. The simulation grid is $1300\times400\times400$, corresponding to about 208 million grid points, and we maintain a grid spacing of about 0.0025D in the vicinity of the ellipsoid to resolve the boundary layer with 4-5 grid points.

We set the timestep size, $\Delta t = 0.0005$ to keep the CFL number about 0.2 which is necessary to keep the simulation stable using the Adams-Bashforth convection scheme. The hardware used was a single NVIDIA A100 node with 4 GPUs connected with PCIe. The simulation simulation took 50 hours to simulate 20,000 timestep to generate the reported data till $tU_\infty /L = 10$. We observed, on average, a wall time of 0.019 secs per iteration in the pressure Poisson solver, including communication (which consumed about 2\% of the iteration time), as compared to 0.41s with 1 node of an Intel Xeon CPU (2 CPU devices per node). The flow visualization using the volume rendering of vorticity magnitude is shown in figure \ref{fig:ell_array}(b) and the corresponding quantitative metric as drag coefficient for each ellipsoid is shown in figure \ref{fig:ell_array}(c).

\section{Summary}

This paper presents a methodology for GPU-acceleration of the sharp-interface immersed boundary method on NVIDIA GPUs to perform computational fluid dynamics simulations on a Cartesian grid around complex geometries. Our GPU native method consists of three key featuresfor GPU porting: (a) The advection-diffusion equation solver, (b) Treatment of Immersed Body, and (c) Pressure Poisson solver. With the current focus on stationary bodies, we identify the Poisson pressure solver as the most time-consuming module, especially due to the communication across multiple GPUs. In this work, we use MPI non-blocking communication directives to optimize communication and achieved over 90\% weak and strong scaling efficiencies. Further, we verified the code using two test cases to assess accuracy and speedup relative to the CPU version of the already-validated code, ViCar3D. The test cases were flow over a 2D cylinder and flow over a rectangular wing at an angle-of-attack of $25^\circ$. We estimate the GPU acceleration to achieve over $20\times$ speedup compared to the CPU code over two GPU types - A100 and L40s, with each set of GPUs installed on a single node. These tests validated the success of the GPU porting and provided a quantitative measure of the acceleration achieved. Next, we challenged the code to perform CFD simulations around complex-shaped objects and multi-object arrangements. Both would challenge body-conformal grid methods but were extremely trivial on a Cartesian grid and fast with a GPU, even on a single node with NVIDIA A100 GPUs. 

The current demonstration of GPU-acceleration has resulted in significant speedups in the simulation of flow around stationary bodies. In future work, our focus will be on two areas: demonstrating simulations of moving body problems with the GPU code and efficient scaling of the code over multi-node GPU systems. Moving-body CFD simulations present a different challenge compared to stationary-body simulation, especially regarding GPU-accelerated immersed-body handling subroutines. Next, scaling the code across multiple nodes requires advanced topology-aware communication, as GPUs on different nodes are connected via slower interconnects compared to faster intra-node interconnects, such as NVlink within a node. The work on the following two areas will result in a complete GPU-accelerated procedure for simulating fluid flow around complex moving bodies and fluid-structure interaction.

\section*{Acknowledgments}
The development of the GPU solver benefited from NSF grant - CBET-
2011619 and ONR grants - N00014-22-1-2655 and N00014-24-12516. The work benefited from the computational resources provided through the DoD HPC Modernization Program. 

\bibliography{sample}

@book{ruetsch2024cuda,
  title={CUDA Fortran for scientists and engineers: best practices for efficient CUDA Fortran programming},
  author={Ruetsch, Gregory and Fatica, Massimiliano},
  year={2024},
  publisher={Elsevier}
}

@article{mittal2008versatile,
  title={A versatile sharp interface immersed boundary method for incompressible flows with complex boundaries},
  author={Mittal, Rajat and Dong, Haibo and Bozkurttas, Meliha and Najjar, FM and Vargas, Abel and Von Loebbecke, Alfred},
  journal={Journal of computational physics},
  volume={227},
  number={10},
  pages={4825--4852},
  year={2008},
  publisher={Elsevier}
}

@article{menon2022contribution,
  title={Contribution of spanwise and cross-span vortices to the lift generation of low-aspect-ratio wings: Insights from force partitioning},
  author={Menon, Karthik and Kumar, Sushrut and Mittal, Rajat},
  journal={Physical Review Fluids},
  volume={7},
  number={11},
  pages={114102},
  year={2022},
  publisher={APS}
}

@article{kumar2025computational,
  title={Computational modelling and analysis of the coupled aero-structural dynamics in bat-inspired wings},
  author={Kumar, Sushrut and Seo, Jung-Hee and Mittal, Rajat},
  journal={Journal of Fluid Mechanics},
  volume={1010},
  pages={A53},
  year={2025},
  publisher={Cambridge University Press}
}

@article{mittal2023origin,
  title={Origin and evolution of immersed boundary methods in computational fluid dynamics},
  author={Mittal, Rajat and Seo, Jung Hee},
  journal={Physical review fluids},
  volume={8},
  number={10},
  pages={100501},
  year={2023},
  publisher={APS}
}

@article{mittal2025freeman,
  title={Freeman Scholar Lecture (2021)—Sharp-Interface Immersed Boundary Methods in Fluid Dynamics},
  author={Mittal, Rajat and Seo, Jung-Hee and Turner, Jacob and Kumar, Sushrut and Prakhar, Suryansh and Zhou, Ji},
  journal={Journal of Fluids Engineering},
  volume={147},
  number={3},
  year={2025},
  publisher={American Society of Mechanical Engineers Digital Collection}
}

@article{viola2023gpu,
  title={GPU accelerated digital twins of the human heart open new routes for cardiovascular research},
  author={Viola, Francesco and Del Corso, Giulio and De Paulis, Ruggero and Verzicco, Roberto},
  journal={Scientific reports},
  volume={13},
  number={1},
  pages={8230},
  year={2023},
  publisher={Nature Publishing Group UK London}
}

@article{viola2022fsei,
  title={FSEI-GPU: GPU accelerated simulations of the fluid--structure--electrophysiology interaction in the left heart},
  author={Viola, Francesco and Spandan, Vamsi and Meschini, Valentina and Romero, Joshua and Fatica, Massimiliano and de Tullio, Marco D and Verzicco, Roberto},
  journal={Computer physics communications},
  volume={273},
  pages={108248},
  year={2022},
  publisher={Elsevier}
}

@article{walker1996mpi,
  title={MPI: a standard message passing interface},
  author={Walker, David W and Dongarra, Jack J},
  journal={Supercomputer},
  volume={12},
  pages={56--68},
  year={1996},
  publisher={ASFRA BV}
}

@article{raj2023gpu,
  title={A GPU-accelerated sharp interface immersed boundary method for versatile geometries},
  author={Raj, Apurva and Khan, Piru Mohan and Alam, Md Irshad and Prakash, Akshay and Roy, Somnath},
  journal={Journal of Computational Physics},
  volume={478},
  pages={111985},
  year={2023},
  publisher={Elsevier}
}

@inproceedings{lopez2014verification,
  title={Verification and Validation of HiFiLES: a High-Order LES unstructured solver on multi-GPU platforms},
  author={L{\'o}pez, Manuel R and Sheshadri, Abhishek and Bull, Jonathan R and Economon, Thomas D and Romero, Joshua and Watkins, Jerry E and Williams, David M and Palacios, Francisco and Jameson, Antony and Manosalvas, David E},
  booktitle={32nd AIAA applied aerodynamics conference},
  pages={3168},
  year={2014}
}

@inproceedings{watkins2016multi,
  title={Multi-GPU, implicit time stepping for high-order methods on unstructured grids},
  author={Watkins, Jerry E and Romero, Joshua and Jameson, Antony},
  booktitle={46th AIAA Fluid Dynamics Conference},
  pages={3965},
  year={2016}
}

@article{kumar2025mechanical,
  title={Mechanical Intelligence in Propulsion via Flexible Caudal Fins},
  author={Kumar, Sushrut and McHenry, Matthew J and Seo, Jung-Hee and Mittal, Rajat},
  journal={arXiv preprint arXiv:2503.23652},
  year={2025}
}

@article{CHORIN196712,
title = {A numerical method for solving incompressible viscous flow problems},
journal = {Journal of Computational Physics},
volume = {2},
number = {1},
pages = {12-26},
year = {1967},
issn = {0021-9991},
doi = {https://doi.org/10.1016/0021-9991(67)90037-X},
url = {https://www.sciencedirect.com/science/article/pii/002199916790037X},
author = {Alexandre Joel Chorin}
}

@article{YANG2014695,
title = {Acceleration of the Jacobi iterative method by factors exceeding 100 using scheduled relaxation},
journal = {Journal of Computational Physics},
volume = {274},
pages = {695-708},
year = {2014},
issn = {0021-9991},
doi = {https://doi.org/10.1016/j.jcp.2014.06.010},
url = {https://www.sciencedirect.com/science/article/pii/S0021999114004173},
author = {Xiyang I.A. Yang and Rajat Mittal}
}

\end{document}